\documentclass[
reprint,
nofootinbib,
amsmath,amssymb,amsbsy,
prd,
onecolumn,
superscriptaddress,
longbibliography,
preprintnumbers]{revtex4-1}

\usepackage[T1]{fontenc}
\usepackage{gfsartemisia-euler}
\usepackage{enumitem}
\usepackage{pifont}
\usepackage{amsmath}

\usepackage[mathscr]{euscript}

\makeatletter
\newcommand{\subalign}[1]{%
  \vcenter{%
    \Let@ \restore@math@cr \default@tag
    \baselineskip\fontdimen10 \scriptfont\tw@
    \advance\baselineskip\fontdimen12 \scriptfont\tw@
    \lineskip\thr@@\fontdimen8 \scriptfont\thr@@
    \lineskiplimit\lineskip
    \ialign{\hfil$\m@th\scriptstyle##$&$\m@th\scriptstyle{}##$\hfil\crcr
      #1\crcr
    }%
  }%
}
\makeatother

 \usepackage{mathtools}
 \usepackage{lmodern}
 \usepackage{lmodern}
\usepackage{graphicx}
\usepackage{hyperref}
\usepackage{amsfonts}
\usepackage{tocvsec2}
\usepackage{slashed}
\usepackage{cancel}
\usepackage{comment}

\newcommand\beq{\begin{eqnarray}}
\newcommand\eeq{\end{eqnarray}}

\newcommand{\half}{\frac{1}{2}}

\newcommand\eqn[1]{\label{eq:#1}} 
\newcommand\eq[1]{eq.~(\ref{eq:#1})}

\newcommand{\bfx}{{\mathbf x}}

\newcommand{\CD}{{\cal D}}

\newcommand{\CL}{{\cal L}}

\newcommand{\mybar}[1]{\kern 0.6pt\overline{\kern -0.6pt#1\kern -0.6pt}\kern 0.6pt}

\def\half{\tfrac{1}{2}}

\def\U\Omega{U(1)_{\Omega}}

\hypersetup{
  colorlinks=true,   
  linkcolor=cyan,    
  citecolor=blue,    
  filecolor=magenta, 
  urlcolor=blue      
}
\usepackage[section]{placeins}

\newcommand{\ddx}[1]{\frac{\partial}{\partial  #1}}

\newcommand{\si}{\,\;}
\newcommand{\rmd}{{\rm d}}
\begin{document}
\title{Chiral edge states on spheres for lattice domain wall fermions}

\preprint{INT-PUB-24-057}

\author{Michael Clancy}
\email{mclancy2@uw.edu}
\affiliation{Institute for Nuclear Theory, Box 351550, Seattle, WA 98195-1550, USA}
\author{David B. Kaplan}
\email{dbkaplan@uw.edu}
\affiliation{Institute for Nuclear Theory, Box 351550, Seattle, WA 98195-1550, USA}


\begin{abstract}
Recently Weyl edge states on manifolds in  dimension $d+1$ with a connected  $d$-dimensional boundary were proposed as candidates for lattice regularization of chiral gauge theories, for even $d$.  The examples considered to date include solid cylinders in any odd dimension, and the 3-ball with boundary $S^2$.  Here we consider the general case of a $(d+1)$-dimensional  ball for any even $d$ and show that the  theory for the edge states on $S^d$ describe a conventional Weyl fermion on a sphere with half-integer  momenta.  A possible advantage of such theories is that they can be discretized by a square lattice without breaking the underlying discrete hypercubic symmetry.
 
\end{abstract}
\maketitle


\section{Introduction}
In Ref.~\cite{kaplan2024chiral} it was pointed out that manifolds with a single boundary  host Weyl fermion edge states and could provide the paradigm for a lattice regulator for chiral gauge theories.  In particular, it was argued that the Weyl edge states must also occur for Wilson fermions on a spacetime lattice, despite the Nielsen-Ninomiya theorem, given that their existence is mandated at the boundary between topological phases, and that nontrivial topological phase structure has been found for Wilson fermions \cite{Golterman:1992ub}. The existence of such Weyl edge states on the lattice was shown in a simple lattice calculation in Ref.~\cite{kaplan2024weyl};   related work may also be found in Refs.~\cite{aoki2022curved,aoki2023curved,aoki2024,aoki2024study}. 

 Ref.~\cite{kaplan2024chiral} considered a cylindrical geometry, and the point of this paper is to derive similar results for spherical geometry.    The case of edge states on $S^2$  was analyzed previously in Ref.~\cite{aoki2022curved,aoki2023curved,aoki2024,aoki2024study}, and our results agree with those references; the main practical value of the present work is the generalization to edge states on $S^4$.    We show that the edge states have the conventional action for a free fermion on a sphere, how the spin connection arises from the domain wall formulation, and how to see that the momenta of the boundary fermion takes half-integer values -- as seen already for the case of fermions on the surface of a solid cylinder \cite{kaplan2024chiral}.  We do not consider a lattice version of the theory, but note that because of the geometry's spherical symmetry, a discretized version on a square lattice will preserve the entire hypercubic symmetry group, unlike the latticized cylinder.  This enhanced symmetry could prove useful in attaining the continuum limit of such a theory.

A key part of the proposal in Ref.~\cite{kaplan2024chiral} dealt with the problem of how to gauge the theory such that it looks like a local, $d$-dimensional gauge theory and not $(d+1)$-dimensional. The proposed method  is to have dynamical gauge fields solely on the boundary, while bulk gauge fields were deterministically derived from the quantum fields as a boundary condition, as introduced in Ref.~\cite{grabowska2016nonperturbative}.
Whether this procedure can work has been called into question on symmetry grounds in Ref.~\cite{golterman2024conserved}, which concerns are likely related to the existence of bulk zeromodes discussed previously in  Ref.~\cite{aoki2023curved,aoki2024,aoki2024study}.  These concerns will be addressed elsewhere; in this paper, however, we do not consider gauge fields, but simply discuss the properties of free fermions.


\section{Domain wall on the cylinder}
We begin with a brief review of the cylinder geometry analyzed in \cite{kaplan2024weyl}.  In order to obtain a theory of a Weyl fermion in $d$ dimensions, we start with a Dirac fermion $\psi$ in $d+1$ Euclidian dimensions with cylindrical symmetry.  The action is
\beq \eqn{diskbulkaction}
S = \int d^{d-1} x_\parallel\, d^2 x_\perp \, \overline{\psi} \CD \psi, \quad \CD = \slashed{\partial}_\parallel + \slashed{\partial}_\perp + m(r).  
\eeq
Here $\bfx_\parallel$ are coordinates for the directions along the length of the cylinder and can be ignored for our purposes, while $\bfx_\perp$ are coordinates for the cross-sectional disk, which can be chosen to be the Cartesian coordinates $\{x,y\}$ or polar coordinates $\{r,\phi\}$.  The $r$-dependent mass $m(r)$ is taken to have the profile
\beq
m(r) = \begin{cases} m & r<R\\ -M & r>R \end{cases}\ ,\qquad r=\sqrt{x^2+y^2}\ ,
\eqn{mass}
\eeq 
with $m$ and $M$ positive. In the Cartesian basis we take coordinate independent $\gamma$-matrices satisfying the usual Clifford algebra.  We also define the coordinate-dependent $\gamma^r$ and $\gamma^\phi$ matrices
\beq
\gamma^r = \hat r\cdot\vec\gamma\ ,\qquad
\gamma^\phi = \hat{\phi} \cdot\vec\gamma\ ,
\eeq
where $\hat r$ and $\hat \phi$ are the radial and azimuthal unit vectors in the plane of the cross-sectional disk.  They too obey the Clifford algebra. 
In terms of these $\gamma$-matrices the action may be written as
\beq \eqn{diskbulkaction2}
S&=& \int  d^{d-1} x_\parallel\,r dr\, d\phi \, \overline{\psi} \left[ \slashed\partial_\parallel+\gamma^r\ddx{r} + \frac{1}{r}\gamma^\phi \ddx{\phi} \right] \psi
\cr &&\cr
&=& \int  d^{d-1} x_\parallel\,r dr\, d\phi\, \overline{\psi} \left[ \slashed\partial_\parallel+\gamma^r \left(\ddx{r} + \frac{1}{2r} \right) + \frac{i}{r}\gamma^\phi  J   + m(r)\right] \psi\ ,
\eqn{act2}
\eeq
where $ J $ is the total angular momentum operator,
\beq
 J \equiv \left(-i\ddx{\phi} + \frac{1}{2} \Sigma \right)\ ,\qquad \frac{1}{2}\Sigma =- \frac{i}{4}\left[ \gamma^x,\gamma^y\right] =  -\frac{i}{4}\left[ \gamma^r,\gamma^\phi\right]\ .
\eeq
As discussed in Ref.~\cite{kaplan2024chiral}, the $\psi$ and $\bar\psi$ are conveniently expanded in eigenfunctions of $\bar\CD \CD$ and $\CD\bar\CD$ respectively, where $\bar\CD $ is the adjoint of $\CD$ with respect to the integration measure.  These eigenfunctions are  represented in terms of spinor harmonics, which for   the cylindrical geometry consist of  two-component spinors eigenstates of the total angular momentum $ J $, the two components corresponding to orbital angular momenta $\ell$ and $\ell+1$.  The eigenvalue $\ell$ takes integer values, reflecting the fact the eigenfunctions are single-valued on the cylinder.    The branch of solutions of particular interest are the chiral edge states, which are localized at $r=R$ and are not gapped in the limit of large $R$.  In the limit of large $m,M$, the  edge state is found to have the form 
\footnote{In a lattice simulation one effectively takes $M\to\infty$ and works with a finite volume lattice.  On the other hand, one must take $m = O(1/a)$, where $a$ is the lattice spacing, in order to simulate a nontrivial topological phase with massless edge states. In the continuum limit of the lattice system, however, one only considers edge states with momenta much less then $1/a$ and so the large-$m$ behavior discussed here pertains.  }
\beq
\psi_\text{edge}= f(r) \,\chi(\bfx_\parallel,\phi)\ ,\qquad \frac{1+\gamma^r}{2}\chi = 0\ ,
\eeq
where the profile function $f(r)$ satisfies
\beq
\left[-\left(\ddx{r} + \frac{1}{2r} \right) + m(r)\right]f(r) = 0\ .
\eeq
In the limit $m,M\to\infty$ this state becomes exactly localized at the boundary $r=R$, and the effective action for the edge state can be found by dropping the gapped modes and performing the $r$ integration in \eq{act2}, effectively just removing the $\gamma^r \left(\ddx{r} + \frac{1}{2r} \right)+ m(r)$ terms from the action,  obtaining
\beq \eqn{diskedgeaction}
S_\text{edge} = \int  d^{d-1} x_\parallel\, d\phi \,\overline{\chi}\bar P_-\left(\slashed\partial_\parallel+\,\frac{i}{R}\gamma^\phi  J  \right)\bar P_+\chi.
\eeq
where we have written the Weyl fermion as a Dirac fermion subject to the projection operators
\beq
\bar P_\pm = \frac{1\pm\gamma^r}{2}\ ,
\eeq

This action does not na\"ively look like the conventional action for a free Weyl fermion propagating on the flat $d$-dimensional surface of a cylinder, even with the substitution of $\phi$ for the coordinate $x_d = R\phi$, due to the presence of the spin $\Sigma$ in the action, as well as the fact that the $\bar \gamma^r$ and $\gamma^\phi$ matrices depend on $x_d$.  Furthermore,  while the eigenfunctions of $\bar\CD \CD$ and $\CD\bar\CD$ have integer orbital angular momentum, and hence $\chi(\bfx_\parallel,\phi) $ is periodic in $\phi$, the eigenvalues of $ J $ are $j=\ell\pm 1/2$ for integer $\ell$, and so the edge state behaves like a free particle on the circular boundary quantized with anti-periodic boundary conditions. 
This situation can be easily rectified by a change of variables, 
\beq
\chi= V \chi'\ ,\qquad \bar\chi = \bar\chi' V^\dagger\ ,\qquad V = e^{-i \phi \Sigma/2}\ ,
\eqn{cov}\eeq
thanks to the identities
\beq
V^\dagger\gamma^r V = \gamma^x\ ,\qquad V^\dagger\gamma^\phi V = \gamma^y\ ,\qquad  V^\dagger  J  V =V^\dagger \left(-i \partial_\phi + \half\Sigma\right)V =-i\partial_\phi\ .
\eeq
In the new basis the boundary action becomes that of a conventional right-handed Weyl fermion on $S^1\times R^{d-1}$, namely
\beq \eqn{diskedgeactionstraight}
S_\text{edge} = \int  d^{d} x   \,\overline{\chi} P_L\slashed\partial\, P_R\chi.
\eeq
where $x_d$ is the coordinate on a circle of circumference $2\pi R$. The $d$ $\gamma$-matrices are now all coordinate independent and consist of $\vec\gamma = \{\vec\gamma_\parallel,\gamma^y\} $, while the projection operators are
\beq
P_{L,R} = \frac{1\pm\Gamma_5}{2}\ ,\qquad \Gamma_5 \equiv \gamma^x\ .
\eeq
Finally, the half-integer values for the momentum component $p_d$ are evident because $V$ is double-valued on $S^1$, so that while $\chi$ was originally single-valued in $\phi$, after the change of variables in \eq{cov} we have $\chi(\bfx_\parallel,\phi) = - \chi(\bfx_\parallel,\phi+2\pi)$.  Thus the half-integer momenta arise because $\chi$ in the new basis obeys anti-periodic boundary conditions on the circle.

We will see that a similar analysis can be adopted for Weyl edge states on spherical boundaries as well, although it requires a bit more work to see this due to  the nonzero curvature of the sphere.


\section{Weyl edge states on spherical boundaries}


 \subsection{Domain wall fermions with spherical symmetry}
 

The Lagrange density for a $(d+1)$-dimensional theory of Dirac fermions with massless Weyl fermion edge states bound to a spherical $S^d$ manifold may be written as
\beq
\CL= \bar\psi \CD \psi\ ,\qquad \CD = \slashed{\partial} + m(r)\ ,
\eqn{DiracS}
\eeq
where  we take $m(r)$ to be the radially-dependent mass
\beq
m(r) = \begin{cases} m & r<R\\ -M & r>R \end{cases}\ ,
\eqn{massS}
\eeq
 with $r=|\bfx|$.

We can rewrite $\slashed \partial$ in terms of the  rotation generators,
\beq J _{ab} = L_{ab} -\frac{i}{4}\left[\gamma_a,\gamma_b\right],\qquad L_{ab} = -i\left(x_a \partial_b - x_b \partial_a\right)\ ,
\eeq
using the identity
\beq
 \frac{i}{r} \gamma_b\hat r_a \left( \frac{i}{4}\left[\gamma_a,\gamma_b\right]\right) = 
 \frac{d}{2r}\gamma_r\ ,
 \qquad \hat r = \frac{\bfx}{|\bfx|}\ ,\qquad \gamma_r \equiv \hat r\cdot \vec\gamma\ ,
 \eeq 
 with the result
 \beq
\slashed{\partial} = \gamma_r \left(\partial_r + \frac{d}{2r}\right) + \frac{i}{r} \gamma_b\hat r_a  J _{ab}\ .
\eeq
As before, the tilde on $\gamma_r$ signifies a coordinate-dependent matrix, while the $\gamma_a$ (without a tilde) are constant.

In  analogy with the cylinder example, one can expand $\psi$ and $\bar\psi$ in eigenstates of $\bar \CD \CD$  and $\CD\bar\CD$ respectively, where  $\bar\CD$ is the adjoint of $\CD$ with respect to the integration measure.  Again the eigenfunctions are spinor harmonics which are single-valued on $R^{d+1}$.  Of particular interest are the chiral edge states which form the only branch of solutions that is gapless in the $R\to\infty$ limit.
In the limit of large $m$ and $M$ these edge state eigenfunctions take the form
\beq
\psi = f(r) \chi({\boldsymbol \Omega})\ ,\qquad \bar P_+\chi = 0\ ,\qquad \bar P_\pm = \frac{1\pm \bar \gamma_r}{2}\ .
\eeq
where ${\boldsymbol \Omega}$ are the $d$ angles parametrizing the   spherical mass defect and $\chi$ is chiral spinor which is  single-valued in ${\boldsymbol \Omega}$.  For large $M$ and $ m$   the profile function $f(r)$ satisfies
\beq
\left[-\left(\partial_r + \frac{d}{2r}\right)+ m(r)\right]f(r) = 0\ ,
\eeq
becoming completely localized at the mass defect as $M$ and $m$ become infinite.
The effective action for the edge state can be written in this limit as $d$-dimensional action
\beq \eqn{sphereedgeaction}
S_\text{edge} = R^d\int  d^{d} {\boldsymbol\Omega} \ \overline{\chi}\bar P_+\left(\frac{i}{R} \gamma_b \hat r_a J_{ab}\right)\bar P_-\chi\ ,
\eqn{dwfS}\eeq
the spherical analog of \eq{diskedgeaction}.


\subsection{Relating the domain wall action to the vielbein formalism}


Although the above expression is rather simple, it does not immediately resemble the   conventional  action one would write for a free fermion on a sphere.  Using the notation of Ref.~\cite{Eguchi:1980jx},  a massless Weyl fermion on a curved manifold has the Lagrange density
\beq \eqn{diracopflat}
\CL_\text{Weyl} =\overline{\chi}P_+\gamma^i   E^\alpha_{\si i} \left(\ddx{ \Omega^\alpha} + \frac{i}{2}   \omega^{jk}_{\ \ \alpha }\Sigma_{jk} \right) P_-\chi\ ,
\eqn{weylS}
\eeq
 where all indices run from $1,\ldots,d$, Greek indices referring to spacetime coordinates  and Latin indices tangent space coordinates. The $\gamma^i$ are coordinate independent Dirac matrices, $  e^i_{\ \alpha}$ is the vielbein, $ E^\alpha_{\ i}$ is its inverse, and $  \omega^{jk}_{\ \ \alpha }$ is the spin connection.     The  $P_\pm$ matrices are the chiral projection operators $(1\pm\gamma_5)/2$, assuming $d$ to be even.  We work in Euclidian spacetime, so the metric for the tangent space is $\eta_{ab} = \delta_{ab}$ and there is no distinction between upper and lower tangent space indices.

   To show the equivalence of the edge state actions in \eq{weylS}  and \eq{dwfS} it is most convenient to rederive the edge state action in a different basis for the fermions, starting again from the $(d+1)$-dimension Lagrangian in \eq{DiracS} rewritten  as
   \beq
   \CL= \overline{\psi} \left[\gamma^a E^\mu_{\si a} \left(\ddx{x^\mu} + \frac{i}{2} \omega^{bc}_{\ \ \mu }\Sigma_{bc} \right) + m(r)\right]\psi\ ,
   \eeq
   where $x^\mu$ are Cartesian coordinates, and so we have
   \beq
   E^\mu_{\si a} = \delta^\mu_{\si a}\ ,\qquad  \omega^{bc}_{\ \ \mu }=0
   \ ,\qquad a,\mu = 0,\ldots,d\ .
   \eeq
   If we now make a coordinate transformation to spherical coordinates, $x^\mu \to \bar x^\mu$, where $\bar x^\mu =\Omega^\mu$ for $\mu=1,\ldots,d$, and $\bar x^\mu =r$ for $\mu=d+1$, where $r$ is the radial coordinate and ${\boldsymbol \Omega}$ are  the angles  on $S^d$.
   Under this coordinate transformation, the vielbein transforms to
   \beq
   e^a_{\ \mu} \to e^a_{\ \nu}\frac{\partial x^\nu}{\partial \bar  x_\mu}\ .
  \eqn{spherical} \eeq
 Since  $(\partial \bar x^\mu/\partial x_\nu)$ is the gradient of the coordinate $\bar x^\mu$, it is proportional to the unit vector pointing in the $\bar x^\mu$ direction, e.g. $\hat\Omega^\mu$ for $\mu=1,\ldots,d$ and $\hat r$ for $\mu=d+1$.  That means $(\partial  x^\nu/\partial \bar x_\mu)$ is a matrix $M^\nu_{\ \mu}$ whose columns are proportional to $\hat n^\mu$ and can be rendered diagonal if we simultaneously perform a  gauge transformation in the tangent space by the orthogonal matrix $\left[O^{-1}\right]^a_{\ b} = (\hat n_a)_b$ (the $b^{th}$ component of the  unit vector pointing in the direction of the $a^{th}$ spherical coordinate), then the new vielbein ${\bar e}^a_{\ \mu}$ will be a diagonal matrix.  So instead of simply performing the coordinate transformation of \eq{spherical}, we include a gauge transformation and define
 \beq
   e^a_{\ \mu} \to {\bar e}^a_{\ \mu}=[O^{-1}]^a_{\ b}e^b_{\ \nu}\frac{\partial  x^\nu}{\partial \bar x_\mu}=[O^{-1}]^a_{\ b}\delta^b_{\ \nu}\frac{\partial  x^\nu}{\partial \bar x_\mu}\ ,\qquad [O^{-1}]^a_{\ b} = (\hat n_a)_b\ ,
   \eeq
In order to effect this transformation we have to perform a gauge transform $V$ on the fermions where
\beq
\psi = V \psi'\ ,\qquad \bar\psi =\bar\psi' V^\dagger\ ,\qquad V^\dagger \gamma^a V = O^a_{\ b} \gamma^b\ .
\eqn{gt}
\eeq
The $V$ matrix represents the same gauge transformation as $O$, just expressed in the spinor representation instead of the vector representation.  We explicitly construct  $V$    in appendix~\ref{sec:appa} but for now simply assume it exists.   The new vielbein $\bar e$ is the simplest one that one would consider writing down in spherical coordinates.  For example, following this procedure for $d=2$ and ordering the coordinates as $\{\bar x^1,\bar x^2,\bar x^3\} =\{\alpha,\theta,r\}$, where $\alpha = (\pi/2-\phi)$ we obtain ${\bar e}^1 =r\,{\rm d}\alpha$, ${\bar e}^2=r\,{\rm d}\theta$, ${\bar e}^3={\rm d}r$, while the new spin connection $\bar \omega^{bc}_{\ \ \mu }$ is the usual one constructed from these vielbeins\footnote{As discussed in Appendix~A, it is useful to replace the conventional azimuthal angle $\phi$ by $\alpha = \pi/2-\phi$ to facilitate  a simply constructed right-handed coordinate system in all dimensions.}.

     The Lagrangian is now given in terms of spherical coordinates as
     \beq
   \CL = \overline{\psi'} \left[\gamma^a \bar E^\mu_{\si a} \left(\ddx{\bar x^\mu} + \frac{i}{2} \bar \omega^{bc}_{\ \ \mu }\Sigma_{bc} \right) + m(r)\right]\psi'\ ,
   \eeq
   where the $\bar E^\mu_{\si a}$ and $\bar \omega^{bc}_{\ \ \mu }$ are the usual vielbein and spin connection one would choose for spherical coordinates on $R^d$, while the $\gamma$-matrices remain the original, coordinate-independent matrices used to write $\slashed{\partial} $ in \eq{DiracS}.  

 It is useful to rewrite this action as
   \beq
   \CL &=& \overline{\psi'} \left[ \gamma^{d+1} \bar E^r_{\si d+1}  \frac{\partial}{\partial r}    + m(r)  +  i\gamma^i \bar E^\beta_{\si i} \omega^{j}_{\ \ \beta }\Sigma_{d+1,j}   + \gamma^i \bar E^\beta_{\si i}  
   \left(\frac{\partial}{\partial \bar x^\beta}+ \frac{i}{2} 
   \bar \omega^{jk}_{\ \ \beta }\Sigma_{jk}  \right)  \right]\psi'
   \cr &&\cr
   &=&
  \overline{\psi'} \left[ \gamma^{d+1} \left(\frac{\partial}{\partial r}   +\frac{d}{2 r} \right) + m(r)  +  \gamma^i \bar E^\beta_{\si i}\left(\frac{\partial}{\partial \bar x^\beta}+ \frac{i}{2} \bar \omega^{jk}_{\ \ \beta }\Sigma_{jk}  \right)  \right]\psi'\ ,
  \eqn{deq}
  \eeq
   where $\beta,i,j,k$ are summed over $1,\ldots,d$, corresponding to the angular coordinates parametrizing $S^d$.  To obtain the latter result we used the identities (justified in appendix~\ref{sec:appB})
   \beq
 \bar \omega^{ab}_{\ \ d+1 }=0\ ,\qquad   i\gamma^i \bar E^\beta_{\si i} \omega^{d+1,j}_{\ \ \beta }\Sigma_{d+1,j}=\frac{d}{2r}\,\gamma^{d+1}\equiv \frac{d}{2r}\,\Gamma_5\ .
 \eqn{tbp}\eeq
In order to use the familiar notation for chirality, we will equate  $\gamma^{d+1}\equiv \Gamma_5$, which anticommutes with $\gamma^i$ for $i=1,\ldots,d$. In this basis we can once again focus on the edge state $\chi'$, which satisfies 
\beq
\left[ - \left(\frac{\partial}{\partial r}   +\frac{d}{2 r} \right) + m(r) \right]\chi'=0\ ,\qquad P_+ \chi' = \frac{1+\Gamma_5}{2} \chi' = 0\ ,
\eqn{edgeeq}
\eeq
 becoming completely localized on $S^d$ at $r=R$ in the limit of  mass $-M\to-\infty$ for $r>R$ and $m\to \infty$ for $r<R$.  Substituting the edge state $\chi'$ for $\psi'$ in \eq{deq}, making use of \eq{edgeeq}, gives us the expected action for the edge state given in \eq{weylS} -- with the exception  that $\chi$ is replaced by $\chi'$, the prime reminding us that we have performed the gauge transformation $V$ on our field.  Since both \eq{dwfS} and \eq{weylS} have been derived from the same starting point, the Dirac action in $R^{d+1}$, it follows that they are equivalent up to the gauge transformation.

Just as in the case of cylindrical geometry described in the previous section,  the gauge transformation matrix $V$ is double valued on the sphere, and so $\chi'$ behaves like a particle quantized with anti-periodic boundary conditions in the azimuthal angle. This excludes  a surface mode with zero momentum,   consistent with the Dirac operator not having a zeromode on the sphere, see for example \cite{clarkson1994eigenvalues,camporesi1996eigenfunctions,bar1996dirac}.

In conclusion, we have shown that Weyl edge states appear naturally on $S^d$ boundaries of $(d+1)$-dimensional balls, and that their properties can be described by conventional low energy actions in terms of single-valued fields for which the analogue of momentum operators on the sphere are proportional to the generators of $SO(d)$.  Alternatively they can be described using a conventional vielbein and spin connection formalism, at the cost of having to use fields that are double-valued on the sphere -- the two formulations being related by a double-valued gauge transformation in the tangent space.  The calculations in this paper should be relevant for understanding the spectrum of a square lattice realization of the $S^d$ domain wall manifold, where the full hypercubic symmetry is preserved by the sphere (unlike the case of cylindrical geometry).  Other discretization approaches that have been explored for spherical manifolds might also be applied to this formulation, such as radial quantization (see Ref.~\cite{ayyar2024operator} and references therein). 

\section*{Acknowledgements}

This research is supported in part by DOE Grant No. DE-FG02-00ER41132.  DBK would like to thank the Instituto de Fisica Te\'orica at the Universidad Aut\'onoma de Madrid for hospitality while this work was completed, and partial support from the Severo Ochoa Program,  grant CEX2020-001007-S, funded by MCIN/AEI/10.13039/501100011033.

\appendix


\section{Construction of gauge transform $V$} 
\label{sec:appa}

Here we give an explicit construction of the gauge transformation $V$ that appears in \eq{gt}.  We first define our spherical coordinate basis on $R^{D}$, with $D=d+1$ as
\beq
\bar x^\mu = \{\alpha,\theta_{1},\ldots,\theta_{D-2},r\}\ ,\qquad \mu = 1,\ldots,D=(d+1)\ ,\qquad \alpha\equiv \frac{\pi}{2} - \phi\ .
\eqn{xtilde}
\eeq
 Here $r$ is the radial coordinate, the $\theta_i$ are polar coordinates with $\theta_i\in [0,\pi]$, and   $\phi$ is the conventional azimuthal angle taking values on $[0,2\pi)$. The definition of $\alpha$ in terms of $\phi$ and the order chosen for the coordinates are dictated by the desire to have a right-handed coordinate system for any $d$, for having $r$ be associated with the $d+1$ coordinate and therefore $\gamma_{d+1}= \Gamma_5$, and for having a definition that is easily iterated from one value of $d$ to the next. 

The corresponding unit vectors in the tangent space $\{\hat \alpha, \hat\theta_1,\ldots,\hat\theta_{D-2},\hat r\}$ are denoted by $\hat n_\mu$ for $\mu = 1,\ldots,D$.  They can be simply constructed iteratively  in the Cartesian basis $\{\hat x_1,\ldots,\hat x_{D}\}$ for increasing dimension starting with $D=2$, where we add the label $D$ to $\hat n_\mu^{(D)}$ in order  to specify the dimension:
\begin{equation} 
\begin{aligned}
D&=2:\qquad &\hat \alpha^{(2)} &=\{\cos \alpha ,-\sin \alpha \}\ ,\\ 
&&\hat r^{(2)}  &=\{\sin \alpha ,\cos \alpha \} \\ \\
D&=3:\qquad &\hat \alpha^{(3)} &=\{\hat \alpha^{(2)} ,0\}\ , \\
&&\hat \theta_1^{(3)} &=\{\cos\theta_{1}\, \hat r^{(2)},-\sin\theta_{1}\}\ , \\
&&\hat r^{(3)} &=\{\sin\theta_1 \hat r^{(2)},\cos\theta_1\}\  \\ \\
D&>3:\qquad &\hat \alpha^{(D)} &=\{\hat\alpha^{(D-1)},0\}\ , \\
&&\hat \theta_{1}^{(D)} &=\{\hat \theta_1^{(D-1)},0\}\   \\
&&\vdots&\qquad\vdots\  \\ 
&&\hat \theta_{D-3}^{(D)} &=\{\hat \theta_{D-3}^{(D-1)},0\}\   \\
&&\hat \theta_{D-2}^{(D)} &=\{\cos\theta_{D-2}\, \hat r^{(D-1)},-\sin\theta_{D-2}\}\ , \\ 
&&\hat r^{(D)} &=\{\sin\theta_{D-2}\, \hat r^{(D-1)},\cos\theta_{D-2} \}\  .
\end{aligned}
\end{equation}

 The gauge transformation of interest is defined in \eq{gt} as
 \beq 
 V^\dagger \gamma^a V = O^a_{\ b} \gamma^b\ ,\qquad [O^{-1}]^a_{\ b} = [\hat n_a]_b\ ,
 \eeq
 where $O$ is the gauge transformation in the vector representation.  Its action is to rotate the basis vectors from the spherical coordinate frame to the Cartesian one:
 \beq
 [O^{-1}\hat n_a]_c  = \delta_{ac}
 \eeq
 so that after the transformation, $\hat\alpha$ has been rotated into the $\hat x_1$ direction,  $\hat\theta_1$ into the $\hat x^2$ direction, etc., and $\hat r$ into the $\hat x_D$ direction.  To construct $V$, the same rotation in the spinor representation, we first represent $O$ as the sequence of $SO(D)$  rotations:
 \beq
 O^{-1} =\left(\,
 \prod_{j=1}^{D-2} e^{-i \theta_{j} J_{j+1,j+2}}\right) \,
  e^{-i\alpha J_{12}}\,
\eeq 
where $J_{a,b}$
are the $SO(D)$   rotation generators
 \beq
 [J_{a b}]_{cd} = -i \left(\delta_{ac} \delta_{bd} - \delta_{ad}\delta_{bc}\right)\ .
 \eeq
 The same gauge transformation in the spinor representation is then given by replacing $J_{a,b}$ by $SO(D)$ generators in the spinor representation, 
 \beq
 V^\dagger =\left(\,
 \prod_{j=1}^{D-2} e^{-i \theta_{j} \Sigma_{j+1,j+2}}\right) \,
  e^{-i\alpha \Sigma_{12}}\,\,\qquad \Sigma_{ab} = -\frac{i}{4}\left[\gamma^a,\gamma^b\right]\ .
  \eeq
For the case of $D=3$ with an $S^2$ boundary our $V^\dagger = e^{-i\alpha \Sigma_{12}}$ agrees with the gauge transformation discussed in Ref.~\cite{aoki2024} up to global rotations, with our angle $\alpha$  related their azimuthal angle $\phi$ by \eq{xtilde}.

The double-valued nature of $V^\dagger$ is evident by looking at the last factor in its construction, $e^{-i\alpha \Sigma_{12}}$, and noting that the eigenvalues of $\Sigma_{12}$ are integer multiples of $1/2$ and so if $\psi$ is single-valued on the sphere with $\psi(\alpha+2\pi) = \psi(\alpha)$, then $\psi' = V\psi$ obeys anti-periodic boundary conditions with $\psi'(\alpha+2\pi) = -\psi'(\alpha)$.  

\section{Derivation of \eq{tbp}}
\label{sec:appB}

The metric $\bar g$ for $R^{D}$ in spherical coordinates defined in \eq{xtilde} with $D=d+1$ may be written in terms of the metric $\hat g$ on the sphere $S^d$ as
\beq
\bar g_{\mu\nu} = \begin{pmatrix}  r^2 \hat g_{ij} & \\ & 1\end{pmatrix}\ ,\qquad \mu,\nu = 1,\ldots,D\ ,\qquad i,j = 1,\ldots,d\ ,
\eeq
so that the vielbein 1-forms are related as
\beq
\bar e^i = r \hat e^{\,i}\ ,\qquad \bar e^D = {\rmd r}\ .
\eqn{erels}
\eeq
The defining equation for the spin connection 1-forms are given by
\beq
{\rmd }\bar e^a = - \bar \omega^{ab}\wedge \bar e^b\ ,\qquad {\rmd }\hat e^{\,a} = - \hat\omega^{ab}\wedge \hat e^{\,b}\ ,
\eeq
which by means of \eq{erels} immediately allows one to solve for $\bar \omega$ in terms of $\hat \omega$ as
\beq
\bar \omega^{iD} = \hat e^{\,i}\ ,\qquad \bar \omega^{ij} = \hat  \omega^{ij}\ .
\eeq
From this one deduces that for $\bar \omega^{ab}_\mu$ one has
$
\bar  \omega^{ab}_D = 0$ and
\beq
\sum_{\beta=1}^d  i \gamma^i \bar E^\beta_{\ i} \bar \omega^{Dj}_\beta \Sigma_{Dj}  =
\sum_{\beta=1}^d  i \gamma^i \left(\frac{1}{ r}\hat E^\beta_{\ i}\right)\hat e^j_\beta \Sigma_{Dj}   
=
 \frac{i }{  r} \gamma^i \Sigma_{Di} 
=\frac{d}{2r}\,\gamma^D= \frac{d}{2r}\,\Gamma_5\ ,
\eeq
proving the assertion in \eq{tbp} with $D=d+1$.

\bibliography{refs}
\end{document}